# Thick amorphous complexion formation and extreme thermal stability in ternary nanocrystalline Cu-Zr-Hf alloys

Charlette M. Grigorian [a], Timothy J. Rupert [a, b, *]

[a] Department of Chemical Engineering and Materials Science, University of California, Irvine, California 92697
[b] Department of Mechanical and Aerospace Engineering, University of California, Irvine, California 92697
[*] Email address: trupert@uci.edu

Building on the recent discovery of tough nanocrystalline Cu-Zr alloys with amorphous intergranular films, this paper investigates ternary nanocrystalline Cu-Zr-Hf alloys with a focus on understanding how alloy composition affects the formation of disordered complexions. Binary Cu-Zr and Cu-Hf alloys with similar initial grain sizes were also fabricated for comparison. The thermal stability of the nanocrystalline alloys was evaluated by annealing at 950 °C (>95% of the solidus temperatures), followed by detailed characterization of the grain boundary structure. All of the ternary alloys exhibited exceptional thermal stability comparable to that of the binary Cu-Zr alloy, and remained nanocrystalline even after two weeks of annealing at this extremely high temperature. Despite carbide formation and growth in these alloys during milling and annealing, the thermal stability of the ternary alloys is mainly attributed to the formation of thick amorphous intergranular films at high temperatures. Our results show that ternary alloy compositions have thicker boundary films compared to the binary alloys with similar global dopant concentrations. While it is not required for amorphous complexion formation, this work shows that having at least three elements present at the interface can lead to thicker grain boundary films, which is expected to maximize the previously reported toughening effect.

1. **Introduction**

Nanocrystalline metals are extremely strong [1, 2], but are commonly plagued by two major problems: (1) poor thermal stability and (2) limited ductility. The characteristically large volume fraction of grain boundaries in nanocrystalline metals has excess energy, providing a large driving force for grain growth [3]. Significant grain growth at temperatures well below those typically required for materials processing (e.g., powder consolidation or sheet-metal forming) has been reported [4, 5], with some studies even reporting coarsening at room temperature [6-8]. In addition, nanocrystalline metals are usually much more brittle than coarser-grained metals [2, 9]. This issue has been traced to the fact that dislocations must be absorbed at grain boundaries, which can eventually lead to crack nucleation at relatively small plastic strains [10, 11]. Ideally, one would like to find a solution that can resolve both of these persistent issues simultaneously.

The thermal stability of nanocrystalline metals can be enhanced by using dopant elements that segregate to grain boundaries to reduce the grain boundary free energy, thereby reducing the driving force for grain growth. Weissmuller first proposed the deliberate segregation of solute atoms to the grain boundaries as a method of improving thermal stability in order to reduce the excess grain boundary energy to zero [12]. More recently, Polyakov et al. showed that a sputtered Hf-Ti alloy retained an average grain size of 50 nm after annealing for 96 h at 800 °C due to the segregation of Ti to the grain boundaries [13]. Similarly, Chen et al. demonstrated that the segregation of C in ball milled Fe-C alloys improved thermal stability and helped to maintain grain sizes under 15 nm after annealing for 1 h at 573 K [14]. These studies are in agreement on the effect of the dopant atoms on thermal stability but provide limited criteria for selecting this chemistry. These criteria were addressed generally by Murdoch and Schuh, who developed an approach for calculating segregation enthalpies of many binary alloy systems, allowing for a general estimation of whether a dopant alloy will segregate to the grain boundary [15]. In addition, stability maps developed by Darling et. al. determine, for a given grain size and temperature, the minimum solute concentration required to reduce grain boundary free energy to zero [16]. While manipulating the chemistry of grain boundaries has led to marked improvements to thermal stability, this phenomenon alone is not likely to significantly affect the ductility of nanocrystalline materials.

A potential solution to both the limited thermal stability and ductility of nanocrystalline metals is to also consider the effects of grain boundary structure. For example, a recent computational study by Pan and Rupert showed that thick, amorphous grain boundary films can significantly delay crack nucleation and propagation at the interface by acting as highly efficient dislocation sinks [10]. These amorphous intergranular films (AIFs) are a type of complexion, which are grain boundary phases that exist in thermodynamic equilibrium with their abutting bulk phases [17]. Although the vast majority of work characterizing complexions has been performed on ceramics [18-24], AIFs have also recently been characterized in metallic W-Ni and Mo-Ni alloys [25, 26]. Studies have shown that AIFs can lead to enhanced diffusion, which can cause phenomena such as solid state activated sintering [20] or abnormal grain growth [27]. The effect of AIFs on nanocrystalline metal properties was first observed in nanocrystalline Cu-Zr alloys created by ball milling, with these materials displaying superior hardness, strength, and ductility [28, 29] compared to undoped nanocrystalline Cu [30, 31]. This Cu-Zr alloy was also reported to retain an average grain size of 54 nm even after a week of annealing at 98% of the solidus temperature. The findings described above demonstrate the capacity for AIFs to both stabilize the microstructure and improve ductility in nanocrystalline metals, which led Schuler and Rupert to develop materials selection criteria for AIF formation [32]. These authors successfully predicted, and then experimentally proved, that a positive grain boundary segregation enthalpy and a negative enthalpy of mixing are sufficient conditions for AIF formation [32]. However, this study was limited to binary alloy pairs. Ternary combinations of elements, although not required, may add an additional benefit, as the formation of a glassy structure is dependent on the critical cooling rate required to prevent crystallization. One important criterion utilized in the discovery of stable bulk metallic glasses is the incorporation of three or more elements, which decreases the probability of finding the type of local atomic arrangement necessary to nucleate a crystalline phase and therefore reduces the critical cooling rate [33, 34]. We therefore hypothesize that the introduction of a second dopant element, making the grain boundary composition ternary, will encourage the formation of AIFs in nanocrystalline alloys. More complex chemical compositions at the boundary may allow for AIF formation at lower temperatures or even give thicker films along the grain boundary [35]. Improved stability of AIFs would mean smaller grains sizes after processing or in service, while AIF formation at lower temperatures would reduce the input energy required during heat treatment. Thicker AIFs should lead to further improvements in the ductility of nanocrystalline alloys, as Pan and Rupert showed that thicker films can absorb more dislocations before cracking [10].

In this work, we investigate AIF formation and thickness distributions in ternary nanocrystalline Cu-Zr-Hf. Ternary alloys were created by ball milling, with their thermal stability and propensity to form thick amorphous intergranular films compared to baseline binary Cu-Zr and Cu-Hf alloys. Hf concentrations in the ternary alloys were varied in order to investigate the importance of global dopant concentration. It was found that all Cu-Zr-Hf alloys retained nanocrystalline grain sizes even after 2 weeks of annealing at 950 °C. The excellent thermal stability of these alloys is attributed to a combination of grain boundary segregation and the formation of AIFs that serve to stabilize their microstructure. Thicker AIFs were found to form in the ternary alloys than in binary alloys with the same dopant concentration, carbide phase volume fraction, and carbide grain size distribution. These samples provide a very clean comparison, where all aspects of the microstructure are the same, with the exception that the ternary alloys have a more complex grain boundary chemistry (i.e., three types of atoms at the interfaces). The results of this study demonstrate that ternary alloys can contain very thick amorphous complexions, which may potentially enable greater ductility based on prior theoretical predictions [10].

**2. Materials and Methods**

To begin, we must select alloying elements and compositions which will allow for dopant segregation to the grain boundaries and possibly transformation into amorphous complexions. As mentioned previously, Schuler and Rupert developed materials selection guidelines for AIF formation which emphasized a positive enthalpy of segregation and a negative enthalpy of mixing [32]. Bulk metallic glass formation guidelines suggest that multicomponent systems with three or more elements, a negative heat of mixing, and an atomic size mismatch greater than 12% encourage the formation of amorphous phases [33]. The thermal stability and mechanical behavior of nanocrystalline Cu has been widely studied [7, 36, 37], and both Zr and Hf dopants have already been shown experimentally to exhibit strong tendencies to segregate to grain boundaries in Cu [29, 32]. Zr and Hf both possess positive enthalpies of segregation and negative enthalpies of mixing with respect to Cu [15, 38] and have atomic radii that are 25% and 24% larger than Cu, respectively. Although Zr and Hf have nearly identical atomic radii, it has been shown in prior studies that the addition of small amounts of Hf in Zr-based bulk metallic glasses enhances the glass forming ability [39, 40]. In addition, binary phase diagrams for Cu-Zr and Cu-Hf both show deep eutectic points at nearly identical temperatures and concentrations [41, 42]. The similarity of the binary phase diagrams allows for a relatively clear comparison of the ternary mixtures with the two binary baseline systems. Although the ternary phase diagram would be helpful for choosing temperatures and compositions, we could not find a reliable source in the literature.

Binary Cu-Zr, binary Cu-Hf, and ternary Cu-Zr-Hf alloys were produced using elemental Cu (Alfa Aesar, 99.99%, -170 + 400 mesh), Zr (Micron Metals, 99.7%, -50 mesh), and Hf (Alfa Aesar, 99.8%, -100 mesh) powders as starting materials. The powders were ball milled in a SPEX SamplePrep 8000M high-energy ball mill using hardened steel milling media for 10 h under inert atmosphere (99.99% pure Ar) in order to force dopant atoms into solid solution with the Cu matrix, and to ensure grain size refinement. A ball-to-powder weight ratio of 10:1 was employed and 1 wt.% stearic acid was added to the powder mixture prior to milling as a process control agent in order to prevent excessive cold welding. While stearic acid was used for all of the samples discussed in detail here, limited additional powders were processed using hexane, methanol, and toluene as alternative process control agents to observe the effect on contamination during milling. Before milling the targeted alloys, Cu powder and 1 wt.% stearic acid was ball milled for 2 hours and then disposed of in order to coat the vial and milling media with a thin layer of Cu, which helps to minimize Fe contamination from the collisions with the vial walls and media. For the binary alloys, we created samples with low (4-5 at.%) and high (10-11 at.%) dopant concentrations. The Zr concentration in the ternary alloys was kept relatively constant at 4-5 at.% in order to make a direct comparison of their thermal stability and complexion formation behavior to binary alloys. The Hf concentration of the ternary alloys was then varied from 1 to 5 at.%. Sample compositions were determined using energy dispersive X-ray spectroscopy (EDS) in a Tescan GAIA3 scanning electron microscope (SEM) and are displayed in Table 1.

The milled powders were then encapsulated under vacuum in SiO$_2$ tubes, annealed at 950 °C, and then rapidly quenched by immediately dropping into water to freeze in the microstructure which is stable at high temperature. Annealing of the samples encourages the diffusion of Zr and Hf atoms from the Cu matrix to Cu grain boundaries, allowing for the formation of the amorphous phase at the grain boundary through a premelting transition. Such an extreme annealing temperature, greater than 95% of solidus temperature for both binary alloys, was chosen in order to investigate the thermal stability of all samples at temperatures where AIFs are able to form. Annealing times of up to two weeks or 168 h were employed in order to probe the long-term stability of the nanocrystalline alloys. The various dopant concentrations in each sample and anneal times are tabulated in Table 1. All samples were analyzed using X-ray diffraction (XRD), with phase identification, volume fraction, and grain size measurements for both Cu and any second phases made via Rietveld analysis [43]. XRD patterns were collected with a Rigaku SmartLab X-ray diffractometer equipped with Cu K$\alpha$ radiation and a 1D D/teX Ultra 250 detector. A LaB$_6$ standard sample was used to calibrate instrumental parameters, including contributions to peak profile (axial divergence, instrumental broadening) and position (goniometer alignments).

Specimens for investigation with the transmission electron microscope (TEM) were prepared by the lift-out method using a Tescan GAIA3 SEM that is also equipped with a focused ion beam (FIB). A final 2 kV polish of all specimens was performed using a Fischione Model 1040 NanoMill in order to remove excess Ga$^+$ ion beam damage or amorphization of the surface that may have been caused by earlier preparation steps. A JEOL JEM-2800 TEM equipped with dual EDS detectors was used to perform detailed microstructural characterization and chemical analysis. Selected area electron diffraction (SAED) patterns, high resolution imaging, and high-angle annular dark field scanning TEM mode were also employed for further phase identification and microstructural characterization. To confirm the validity of grain sizes obtained via Rietveld refinement of XRD scans, the average grain sizes from at least 100 grains were measured from TEM micrographs for each sample that had been annealed for one week (Table 1). In order to observe solute and second phase distribution throughout the sample, chemical analysis was performed with EDS in order to observe dopant and second phase distribution throughout the microstructure. EDS line scans across grain boundaries were performed in order to confirm segregation of both dopant elements to the grain boundaries in each sample. It is important to note that overlapping grains and slight misalignments from a perfect edge-on condition could not be completely ruled out for these EDS measurements, so the data should be interpreted as a qualitative measurement of boundary chemistry. In order to investigate the effects of the addition of a ternary alloy composition on AIF thickness, at least 50 AIFs were imaged from each the Cu-5Zr, Cu-4Hf, and Cu-4Zr-1Hf samples after annealing for one week. Care was taken to ensure that the grain boundaries investigated were in an edge-on condition by using stage tilting and Fresnel fringe imaging.

## 3. Results and Discussion

*3.1 Microstructural Evolution during Annealing*

XRD patterns and the associated phase identification of all as-milled powders are shown in Figure 1. The nanocrystalline metallic phase is face centered cubic (fcc) in all cases. ZrC and HfC phases have also formed during ball milling and are present in the microstructure for all alloys (Figure 1(b)), even before heat treatment. The formation of carbide phases is likely due to the C-containing process control agent used during mechanical alloying. Possible alternative process control agents were explored (methanol, hexane, and toluene), but these all contain C and lead to similar amounts of carbide formation. Small peaks corresponding to the ZrC phase are evident in the diffraction pattern for the Cu-5Zr but are wider and more prominent in Cu-10Zr. Pronounced HfC peaks are present in all Hf-doped samples, with the intensities of the HfC peaks increasing with increasing Hf concentration. The volume fraction and evolution of the carbide precipitates in each sample will be described in more detail later in the paper. A very small amount of $ZrH_2$ phase is present in all of the as-milled specimens containing Zr, although this phase disappears after annealing.

The average grain size of the fcc phase was calculated from XRD and is displayed in Figure 2 as a function of annealing time. Figure 2(a) presents all data, while Figure 2(b) presents a zoomed view of the Cu-Zr and Cu-Zr-Hf alloys. Grain sizes of ball-milled, pure Cu are also included for comparison. For all annealing times, the vast majority of the alloyed samples exhibit superior thermal stability as compared to the pure Cu sample, with the exception of Cu-11Hf after 336 h (i.e., 2 weeks). Grain sizes calculated via XRD pattern refinement are reliable up to roughly 100 nm [44], which is marked by the orange dashed line in Figure 2. The pure Cu sample grows beyond the XRD resolution by the first measurement (1 h), while the binary Cu-Hf alloys all exceed the XRD resolution by the end of the tests. In contrast, both the Cu-Zr and Cu-Zr-Hf alloys remain nanocrystalline with grain sizes in the range of ~50-65 nm after annealing for 2 weeks at 950 °C. The ternary alloys exhibit grain growth trends that closely track the behavior of the Cu-5Zr alloy (their binary analog, since it has a similar amount of Zr). These results represent a significant increase in thermal stability for nanocrystalline alloys as compared to prior studies. For example, Darling et al. investigated the thermal stability of ball-milled Cu-Ta alloys, reporting an average grain size of 167 nm after annealing for 4 h at 97% of the melting temperature for Cu [45]. In addition, a thermal stability study of a ball-milled W-20 at.% Ti alloy reported an average grain size of 24 nm after annealing at 1100 °C for one week [46]. It is important to note, W is an extremely refractory metal and 1100 °C is only ~32% of the melting temperature of pure W. In comparison, all ternary alloys investigated in this study remain well within the nanocrystalline regime even after 2 weeks of annealing at 96-98% of solidus temperature of the binary alloys.

Figures 3(a-d) present bright field TEM micrographs of the Cu-4Hf and Cu-Zr-Hf alloys that were annealed for 1 week, as well as SAED patterns for phase identification. The SAED patterns confirm the presence of HfC phase dispersed throughout the microstructure in all Hf-doped samples, which can be visually observed in the micrographs at the grain boundaries of the larger Cu grains. No intermetallic phases were found in any of the samples in this study. Grain size distributions measured by TEM for each sample annealed for 1 week are shown in Figure 3(e), and the average values are listed in Table 1. The grain size measurements and phase identification from XRD are in agreement with TEM analysis. Abnormal grain growth, which presents as a bimodal distribution of grain size [27, 47, 48], was not observed. The inhibition of abnormal grain growth in these alloys is in stark contrast to many other materials possessing AIFs, particularly in several ceramic systems [27, 49, 50] where the presence of AIFs is associated with rampant abnormal grain growth. Dillon, Harmer, and coworkers reported abnormal grain growth associated with AIF formation in calcia and silica-doped alumina [48, 51]. Abnormal grain growth in alumina without planned dopants due to AIF formation has also been reported to occur, due to exposure to very small amounts of impurities during the sintering process [9]. However, for the alloys in this present study, the presence of AIFs actually serves to stabilize the microstructure. With the exception of Cu-11Hf, which has a very broad distribution of grain sizes from 30 nm up to 250 nm, all alloys have narrow grain size distributions and retain nanocrystalline grain sizes after 1 week of annealing (Figure 3(e)). This finding is in agreement with previously reported results from a study of a binary Cu-Zr alloy [28, 29]. The formation of AIFs has also been achieved

in a variety of other metallic alloy systems, such as Mo-Ni [26], Ni-W [52], and Cu-Hf [32], with no abnormal grain growth reported.

Average carbide particle sizes and volume fractions measured from XRD patterns, are shown in Figure 4(a) and 4(b) respectively. It is evident that the grain size of the carbide precipitates stays relatively constant in each sample during annealing, with any fluctuations being within the measurement error of XRD grain size measurements. The carbide volume fractions shown in Figure 4(b) also stay constant for most samples, except for the Cu-10Zr and Cu-11Hf samples, where a decrease in carbide phase fraction is found after annealing for 1 h. Despite having 1 at. % less dopant, the Cu-4Hf sample has a significantly higher volume fraction of carbides than the Cu-5Zr sample; after one week of annealing, the Cu-4Hf sample had a carbide volume fraction of 0.023, which is more than double the carbide volume fraction of 0.009 found in Cu-5Zr. This provides an explanation for the difference in thermal stability behavior between these two samples, as there are fewer Hf atoms available to segregate to grain boundaries and thermodynamically stabilize the microstructure. In general, the Cu-Hf alloys have higher volume fractions of the carbide phase than the Cu-Zr alloys with similar overall dopant concentration, suggesting that HfC is easier to form than ZrC under similar processing conditions. This observation is supported by the calculated formation enthalpies of ZrC (-207 kJ/mol) and HfC (-226 kJ/mol) [53]. In contrast, the Cu-5Zr and Cu-4Zr-1Hf alloys have nearly identical carbide phase volumes and also the lowest amounts of all samples investigated in this study, meaning that they should have the same number of dopant atoms free to segregate to grain boundaries to stabilize the microstructure. This is highlighted by the nearly identical thermal stability behavior exhibited by these two samples in Figure 2.

HfC phase distribution throughout the microstructure was further investigated using STEM imaging and EDS. Figure 4(c) shows a high angle annular dark field STEM micrograph of a carbide precipitate at a grain boundary in the Cu-4Hf alloy, which are observed throughout the sample microstructure. An EDS line scan across one of these precipitates in Figure 4(d) demonstrates a high Hf concentration, suggesting that these are HfC particles. While smaller precipitates are indeed scattered throughout the microstructure along the grain boundaries of the Cu-rich, fcc phase, we also occasionally find much larger HfC particles that are ~100-200 nm in diameter. These large precipitates are more common in the samples with larger dopant concentrations.

The microstructures of the Cu-5Zr and Cu-4Zr-1Hf samples annealed for one week were subjected to a more rigorous analysis of the carbide phase distribution throughout the samples. These two specimens were chosen due to their very low, and nearly identical, phase volume fractions, as well as their similar thermal stability behavior. A representative micrograph of the carbide phase distribution in Cu-5Zr annealed for one week is shown in Figure 5(a), showing the presence of slightly larger precipitates at the grain boundaries, but also particles scattered in the grain interiors. Figure 5(b) shows that the size distribution of the carbide particles in the Cu-5Zr and Cu-4Zr-1Hf samples, both at the grain boundaries and within the grain interior, are identical between the two samples. The average size of these particles measured from STEM images agrees well with that measured using XRD. The following discussion will focus primarily on the Cu-5Zr and Cu-4Zr-1Hf samples due to their similarity in microstructure and thermal stability, allowing for the clearest possible investigation of the effects of the addition of a second dopant element on the grain boundary structure.

Since all of the alloys studied here do contain second phase carbide precipitates, it is reasonable to ask whether the thermal stability we observe can be attributed to these particles. Grain size stabilization by Zener pinning can be modeled by [54]:

$$d_Z = \beta \frac{D}{f^m} \qquad (2)$$

where $d_Z$ is the estimated grain size of Zener-pinned grains, $D$ is the second phase particle size, $f$ is the volume fraction of second phase particles, and both $\beta$ and $m$ are constants. Hillert [55] suggested that the constants $\beta = 0.44$ and $m = 0.93$ be used. The grain sizes of Zener-pinned grains for all alloys annealed for one week are shown in Figure 5(c). Grain sizes that are at or above the predictions of Equation 2, in the region that is shaded blue, can possibly be explained as a purely kinetic stabilization caused by second phase particles. On the other hand, grain sizes below these values must be either stabilized by predominantly thermodynamic effects (reduction in grain boundary energy) or, most likely, a combination of kinetic and thermodynamic effects. Of the alloys studied here, only the Cu-11Hf sample falls in the blue region. All of the other samples have grain sizes that are much too small for only kinetic stabilization alone. As such, we next move to confirm dopant segregation to grain boundaries in the different alloys, and investigate the effects of a second segregating dopant element on grain boundary structure; specifically, the formation of thicker amorphous intergranular films which can account for the extreme thermal stability demonstrated by these alloys.

*3.2 Dopant Segregation and Amorphous Complexion Formation*

Previous studies by Khalajhedayati and coworkers [28, 29] reported the formation of amorphous intergranular films in nanocrystalline Cu-Zr alloys processed under conditions similar to those used to create the alloys studied here. To study this behavior, grain boundary segregation of the dopants was observed via STEM and EDS, confirming the conditions necessary for AIF formation. Figure 6(a) shows a HAADF STEM micrograph of the microstructure of Cu-4Zr-1Hf, annealed for 1 week. Regions of bright contrast are seen along several grain boundaries, suggesting segregation of higher Z atoms (Zr and/or Hf) to those boundaries. To confirm dopant segregation, EDS line scans were performed across grain boundaries in Cu-4Zr and Cu-4Zr-1Hf samples annealed for one week, shown in Figures 6(b-e). It is apparent that there are significant spikes in dopant concentration above the noise level inside of the crystals at the grain boundaries, supporting the hypothesis that thermodynamic effects from dopant segregation are also working to stabilize the microstructure. Special care was taken to avoid precipitated secondary phases and ensure that all line scans were taken across grain boundaries only. Even though carbides have formed in the samples doped with Hf, as shown in Figure 4, there remains sufficient unreacted Hf in the alloy to segregate to the grain boundaries. Figure 6(e) also demonstrates the variability of dopant segregation behavior that can occur within a given microstructure, where one boundary is co-doped with both Zr and Hf, while another boundary is doped only by Zr. The occurrence of co-doping of grain boundaries is promising for the hypothesis that thicker, more stable AIFs will form in ternary alloys as compared to binary alloys, since it demonstrates that the multiple added elements have a synergistic effect where the segregating dopants attract each other to the interface rather than compete for segregation sites [56]. This behavior is not surprising, as Zr and Hf have very similar chemical properties with nearly identical atomic and ionic radii [57], and exhibit attractive interactions as evidenced by the fact that they are commonly found together in nature in the same ores [38, 57-59].

Following the confirmation of grain boundary segregation in binary and ternary alloys doped with Hf, high resolution TEM was used to investigate the details of grain boundary structure in each alloy. Figure 7 presents representative examples of complexions found in various alloys annealed for 1 week at 950 °C. Figure 7(a) shows an example of a thick AIF in the binary Cu-5Zr sample, where fast Fourier transforms confirm the amorphous structure of the grain boundary region and the crystallinity of the abutting grains. Obvious AIFs were found in all of the samples presented in this work. However, Figure 7 also makes it clear that these AIFs had a variety of thicknesses. Figure 7(d) shows the thickest AIF found in this study, having a thickness of ~9 nm. On the other hand, the boundary from the Cu-5Zr-5Hf sample that is shown in Figure 7(e) is amorphous but relatively thin in comparison. An important consideration to note is that the grain boundary should be edge-on when measuring the thickness of an AIF in order to ensure accuracy of the measurement. Figure 8(a) shows an example of an AIF found in the Cu-5Zr sample annealed for 1 week, with a thickness of 3.4 nm. Images of the complexion at defocus values of -6, -3, 0, 3, and 6 nm are also shown in order to demonstrate an edge-on imaging condition where the thickness of the AIF does not vary, and guarantees integrity of the measurement. Figure 8(b) shows a similar image series for an AIF in the Cu-4Zr-1Hf sample.

To allow for a meaningful discussion of AIF thickness for different samples, the thicknesses of at least 50 grain boundary complexions from each of the Cu-5Zr, Cu-4Hf, and Cu-4Zr-1Hf alloys annealed for one week from HRTEM micrographs of grain boundaries in an edge-on condition were measured, and are plotted in Figure 9. These three samples were selected for detailed inspection because this allows for the direct observation of the effect of global dopant concentration on AIF thickness, as well as a comparison between binary and ternary alloy compositions with the same global dopant concentration. Additionally, these samples have relatively low carbide volume fractions (all are < 3 vol.%). Since the specimens with a great deal of carbides may have less dopant available for possible segregation, this allows us to make the cleanest possible comparison of grain boundary structure between the samples. It is important to note that the thicknesses of all AIFs from all samples were measured from the thinnest observable region in the edge-on portion of each grain boundary, in order to make conservative measurements and reliable comparisons across all three samples. Taking thickness measurements from the edge-on region of the grain boundary helps to ensure that we do not report artificially high thickness values that come from a boundary being inclined with respect to the viewing direction. An instructive example of the potential error that could be introduced by taking measurements from an improperly aligned region is found in Figure 7(a). Although it may look as though there is a significant variation in thickness across the length of this AIF, only the region of the AIF denoted in the dashed box is in an edge-on condition. In fact, if the region near the top of this figure is inspected at greater magnification, we find evidence of small regions where lattice fringes are visible, suggesting that this is a projection of both a crystal and an amorphous region (i.e., an AIF that is inclined with respect to the viewing direction). Therefore, thicknesses taken from such a region would not be reliable measurements. Figure 8 shows the process by which thickness measurements are reported for each AIF investigated in this study. Measurements are taken from the thinnest observed region of the AIF, with care taken to ensure a perfect edge-on viewing condition. Figure 8(c) shows that the maximum variation in thickness is less than 0.1 nm, meaning that the human error introduced from the thickness measurement is extremely small. The data in Figure 9 is plotted with both a linear scale and a logarithmic scale along the X-axis, to better show the differences in the curves. This data shows that the ternary Cu-4Zr-1Hf alloy is able to form the thickest AIFs, with the cumulative fraction curve furthest to the right in the figure. Analysis of the mean AIF thickness shows that the ternary Cu-4Zr-1Hf alloy (average AIF thickness of 1.85 nm) has AIFs that are on average ~9% thicker than the binary Cu-5Zr alloy (average AIF thickness of 1.7 nm). However, the usage of the mean suppresses some detail, and a more nuanced analysis can be performed on the data set. The cumulative distribution functions of AIF thickness diverge significantly for the thickest ~60% of AIFs observed. The fact that the top half of the AIF thickness curve for the ternary alloy does not overlap at all shows that this material has a population of significantly thicker AIFs. To give another perspective, only 3.9% of AIFs measured in binary Cu-5Zr had thicknesses greater than 4 nm, while 10% of AIFs measured in Cu-4Zr-1Hf had thicknesses greater than 4 nm. It is important to note that this is the cleanest possible comparison among samples in this study, as the only difference between the Cu-5Zr and the Cu-4Zr-1Hf samples is the complexity of the dopant chemistry (the two samples have identical grain size, global dopant concentration, carbide volume fraction, carbide size, and carbide distribution throughout the sample), and a clear difference is found. Although a comparison to the Cu-4Hf sample is not as clean, since this specimen has a higher carbide volume fraction which can reduce the amount of dopant available for grain boundary segregation, we include this data for completeness. There is very little quantitative data about the distribution of AIF thickness available in the literature, so we believe it does add some value to include the raw data.

The advantages of introducing a second dopant element in order to create a ternary alloy composition has important implications regarding grain boundary segregation and the formation of thick, stable AIFs. The bulk metallic glass research clearly shows that a multicomponent composition helps to reduce the free energy penalty associated with the formation of an amorphous phase, and also stabilizes the amorphous phase by lowering the critical cooling rate necessary [33]. In addition, a previous study by Luo et al. has demonstrated that thicker AIFs are able to more easily accept dopant atoms, further increasing the grain boundary dopant concentration and AIF thickness due to this reduction in the free energy penalty [60]. We hypothesize that the ternary alloy composition most directly affects the AIF thickness, which in turn means that more dopant can be accepted by the boundary. This contrasts with a situation where the ternary composition directly leads to higher boundary concentrations, with the higher concentration leading to thicker films. However, a definitive answer to which event happens first or drives the overall behavior is beyond the scope of the present study.

Although we hypothesize that more heavily doped boundaries can sustain thicker AIFs, it is important to remember that alloys with higher global dopant concentrations may also have more precipitation, which can compete with grain boundary segregation. The data regarding the Cu-10Hf alloy in Figure 2 actually show that beyond a certain dopant concentration, the thermal stability of the alloy will be reduced due to the formation of a significant amount of larger carbide particles which do not assist in the pinning of grain boundaries and potentially wick away excess dopant from the fcc phase grain boundaries. It is apparent that in this alloy there is a significant propensity for HfC phase to form, so care must be taken to determine an ideal Hf concentration at which the thick AIFs still play the dominant role in stabilizing the microstructure. Similar behavior was observed in a study by Donaldson et al., in which it was found that annealing of nanocrystalline W-Cr led to the formation of Cr-rich grains similar in size to the W-rich grains, whereas the use of Ti as the dopant led to Ti segregation to grain boundaries upon annealing [11]. The decreased thermal stability of W-Cr as compared to the W-Ti system was attributed to the formation of the large dopant-rich particles throughout the microstructure, which essentially "stole" dopant atoms out of the grain boundaries. In addition to the global dopant concentration, the concentration ratio of dopant elements in ternary alloys is also an important factor that should be considered. Simulations in a prior study by Hu et al. demonstrated a correlation between global dopant concentration and AIF thickness in a binary Al-Cu and a ternary Al-Zr-Cu alloy with a 1:1

Cu:Zr concentration ratio [61]. However, it was also demonstrated that the film thickness could be substantially increased by increasing the ratio of Cu to Zr atoms, while still maintaining the same global dopant concentration.

The Cu-5Zr and Cu-4Zr-1Hf samples investigated in this study provide the cleanest possible comparison of the influence of dopant chemistry on amorphous film thickness. These two samples have the same grain size, carbide volume fraction, carbide size distribution, and spatial distribution of carbides throughout the microstructure, as previously shown in Figures 4(b) and 5(b). The only difference is that there are two types of atoms at the grain boundaries in Cu-5Zr (Cu and Zr atoms), while the boundaries are decorated with three types of atoms in Cu-4Zr-1Hf (Cu, Zr, and Hf atoms). Figure 9 shows that the ternary Cu-4Zr-1Hf alloy has significantly thicker AIFs, which can only be explained by the increased complexity of the grain boundary chemistry.

The formation of AIFs in ternary and multicomponent alloys will have some additional complications compared to similar complexions in binary alloys or single-phase systems. Complicating factors to consider include the potential competition between dopants for grain boundary adsorption sites, as well as attractive or repulsive interactions between different dopant elements [62] , both of which can significantly influence the propensity of an alloy to form AIFs. Previous studies by Zhou and coworkers have investigated the development and experimental validation of grain boundary phase diagrams for binary and multicomponent alloys [63]. These studies show that attractive interactions between dopant elements in the liquid phase led to higher grain boundary excess concentrations, which overpower any effects of site competition and promote grain boundary pre-wetting transitions and disordering at grain boundaries [62, 63]. The evidence of thicker AIFs in the Cu-4Zr-1Hf alloy than the Cu-5Zr alloy possessing the same global dopant concentration supports these findings. The formation of thicker AIFs on average is a sign that the free energy penalty for the formation of an amorphous phase has been reduced by the switch to a ternary grain boundary composition, since AIF thickness is inversely proportional to this energy [60]. The ternary alloy is found to have thicker complexions on average, indicating that the introduction of a second dopant element further stabilizes the amorphous complexion at the grain boundaries. The complexions found in the Cu-Zr-Hf are thicker than any found in metallic systems to date, including Ni-W [52], W-Ni [25], Mo-Ni [26], and Ni-Zr .

Schuler et al. [32] recently showed that AIFs can act to stabilize a nanocrystalline grain structure. AIFs fundamentally form because they have a lower overall energy compared to the "normal" type of grain boundary, which will consequently reduce the driving force for grain growth. In addition, heavy levels of dopant segregation should also kinetically limit grain boundary migration, as these dopants would have to diffuse along with the interface as it moves (or there would be an increase in system energy). The introduction of thicker AIFs to nanocrystalline metals is also desirable due to their demonstrated ability to significantly improve the ductility and fracture toughness of these materials [10, 28]. Pan and Rupert [10] showed that thicker AIFs can absorb more incoming dislocations before cracking, meaning that a polycrystalline sample should be more ductile with thicker AIFs distributed within its microstructure.

## 4. Conclusions

In this study, nanocrystalline Cu-Zr-Hf alloys were created by mechanical alloying in order to gain an understanding of their thermal stability and ability to form amorphous complexions. The microstructures of these alloys were compared to baseline binary Cu-Hf and Cu-Zr alloys to gain insight on the role of each dopant element in stabilizing the microstructure. We draw the following conclusions from the results reported here:

- All ternary alloys investigated in this study showed excellent thermal stability, retaining grain sizes well within the nanocrystalline regime after annealing for two weeks at a temperature above 95% of the solidus temperature for binary Cu-Zr and Cu-Hf alloys. The thermal stability is primarily attributed to the formation of thick amorphous intergranular films, with a minor contribution from HfC and ZrC precipitates located at the grain boundaries.

- Amorphous intergranular films were observed to form in all binary and ternary alloys investigated in this study. Thicker films were found in the ternary Cu-4Hf-1Zr, even when possessing the same global dopant concentration as a binary Cu-5Zr alloy. The introduction of a second dopant element to form a ternary composition in the grain boundary region reduces the energy penalty for AIF formation. Finally, the ternary alloy contained the thickest amorphous complexions observed to date in metallic materials.

The results of this study highlight the enhanced ability to form thick, stable amorphous intergranular films in nanocrystalline alloys by utilizing a ternary composition. Although the grain sizes for the Cu-5Zr and the Cu-Zr-Hf alloys were found to be very similar, the presence of many thick AIFs is advantageous for improving the ductility of a nanocrystalline material. The ability of AIFs to serve as dislocation sinks which can delay crack nucleation and therefore increase fracture toughness is a significant motivator in fabricating nanocrystalline materials for structural applications. The results of this study motivate future work into the investigation of the mechanical properties of ternary alloys with thick AIFs versus binary alloys with thinner, less stable AIFs.

**Acknowledgements**

This study was supported by the U.S. Department of Energy, Office of Basic Energy Sciences, Materials Science and Engineering Division under Award No. DE-SC0014232. SEM, FIB, TEM, and XRD work was performed at the UC Irvine Materials Research Institute (IMRI) using instrumentation funded in part by the National Science Foundation Center for Chemistry at the Space-Time Limit (CHE-0802913).

| Specimen Nominal Composition (at. %) | EDS dopant concentration (at. %) | | XRD grain size (nm) | | | | | TEM grain size (nm) |
|---|---|---|---|---|---|---|---|---|
| | | | Annealing Time | | | | | |
| | Zr | Hf | 0 h | 1 h | 24 h | 168 h | 336 h | 168 h |
| Cu | 0 | 0 | 25 | 108 | 109 | - | 132 | - |
| Cu-5Zr | 4.7 | 0 | 20 | 34 | 44 | 51 | 66 | 50 |
| Cu-4Hf | 0 | 3.9 | 19 | 50 | 59 | 68 | 113 | 63 |
| Cu-10Zr | 9.8 | 0 | 12 | 23 | 24 | 49 | 47 | 39 |
| Cu-11Hf | 0 | 11.3 | 25 | 80 | 90 | 117 | 134 | 103 |
| Cu-4Zr-1Hf | 4.2 | 0.9 | 24 | 44 | 44 | 52 | 62 | 48 |
| Cu-5Zr-3Hf | 4.9 | 2.7 | 21 | 41 | 45 | 45 | 67 | 52 |
| Cu-5Zr-5Hf | 4.8 | 4.7 | 22 | 39 | 54 | 47 | 60 | 50 |

**Table 1.** Chemical compositions, XRD grain sizes, and TEM grain sizes of each alloy investigated in this study. All samples were annealed for 1 h, 24 h (1 day), 168 h (1 week), and 336 h (2 weeks) at 950 °C. The average TEM grain sizes of samples annealed for 168 h (1 week) agree with the XRD grain sizes for the same samples. Dopant concentrations obtained with EDS are typically subject to measurement error of ~1 at.%.

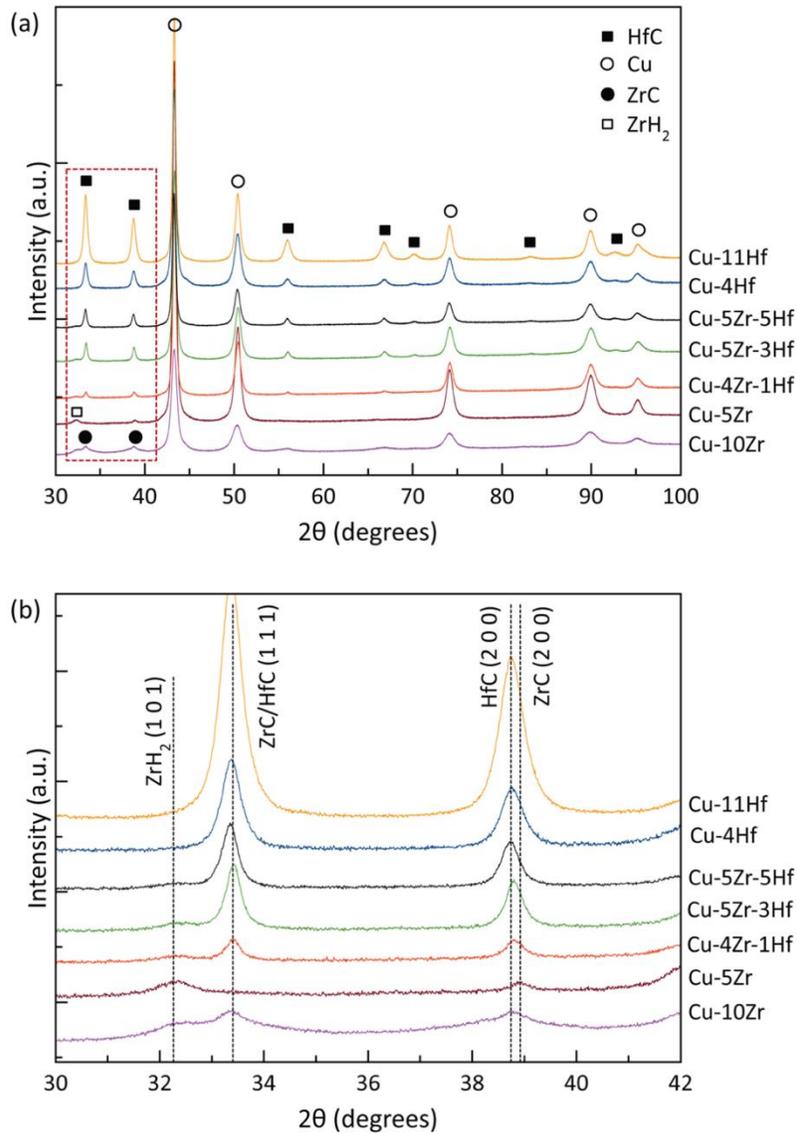

**Figure 1.** (a) X-ray diffraction patterns of the as-milled samples. Phase identification shows the presence of HfC phases for all samples containing Hf, as well as ZrC and ZrH$_2$ phases in Zr-containing samples. (b) Zoomed view of the low angle region where carbide and hydride phases can be seen.

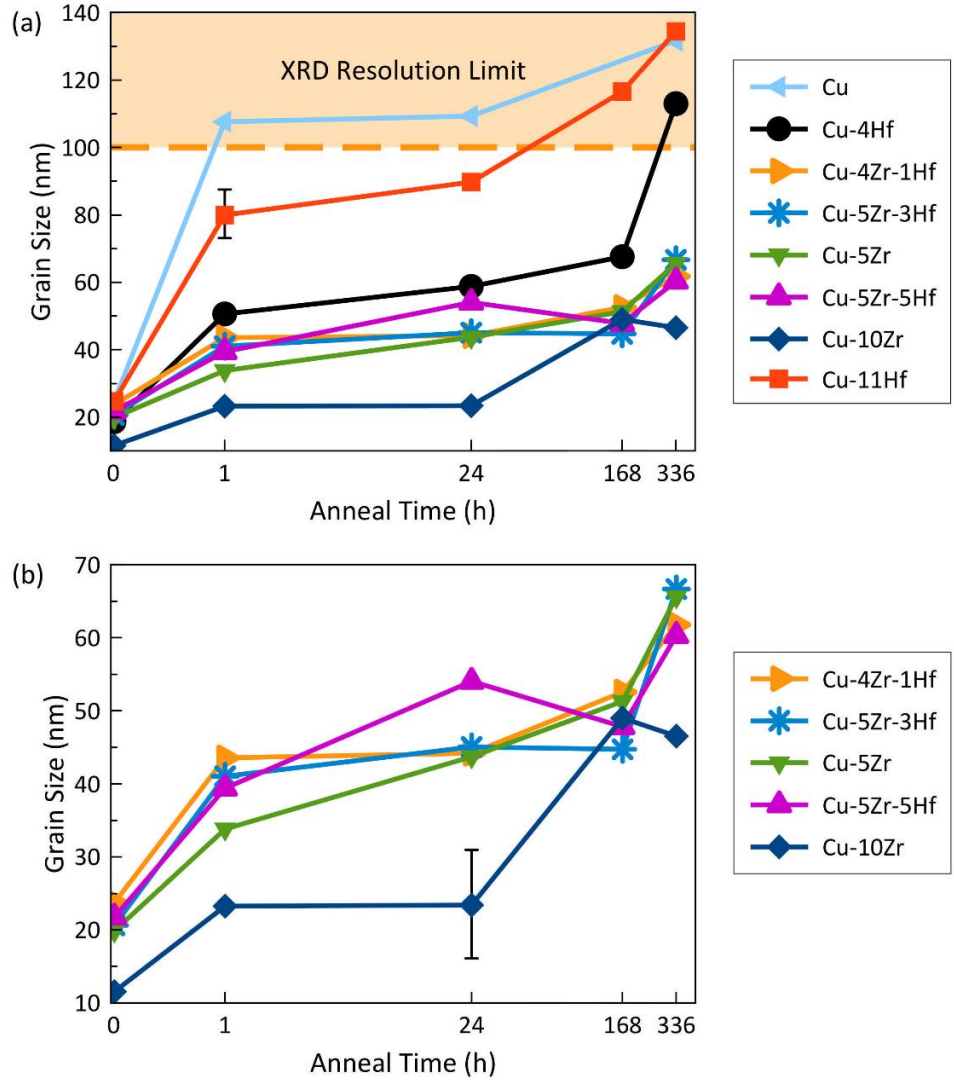

**Figure 2.** (a) Grain sizes for all alloys calculated from XRD patterns as a function of anneal time. Binary Cu-Zr and ternary Cu-Zr-Hf alloys display greater thermal stability than the Cu-Hf alloys. (b) Zoomed view of the grain size data for the Cu-Zr and Cu-Zr-Hf alloys. The 15 nm error bar shown in (a) represents the error in XRD grain size from the results of Rietveld refinement.

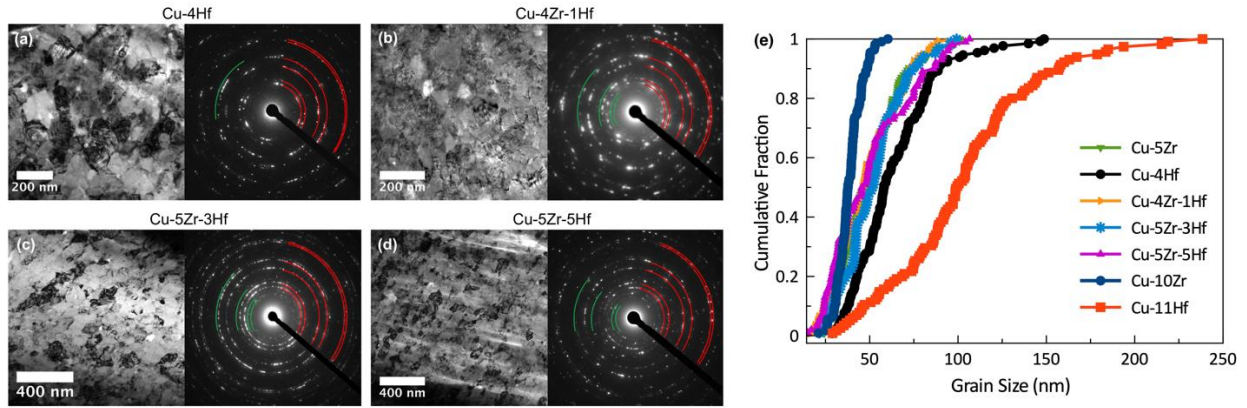

**Figure 3.** Bright field TEM images and SAED patterns for (a) Cu-4Hf, (b) Cu-4Zr-1Hf, (c) Cu-5Zr-3Hf, and (d) Cu-5Zr-5Hf samples that have been annealed for 1 week at 950 °C. SAED patterns confirm the XRD phase identification results, showing the presence of HfC particles. Cu and HfC phases in the SAED patterns are represented by red and green lines, respectively. Grain size distributions after 1 week of annealing at 950 °C are shown in (e). At least 100 grains were measured from TEM micrographs of each sample. A narrow grain size distribution and an average grain size in the nanocrystalline range (<100 nm) is observed for the majority of the samples, with the Cu-11Hf sample being the exception.

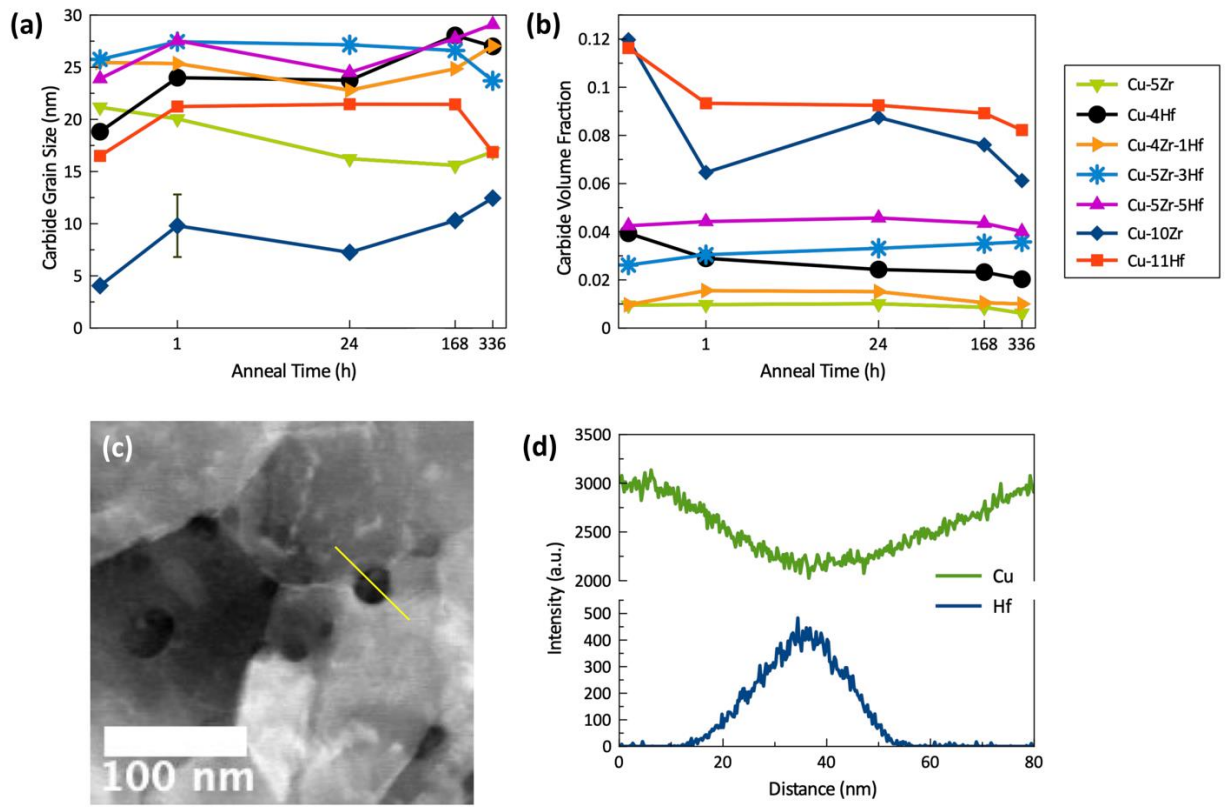

**Figure 4.** (a) Carbide grain size and (b) volume fraction from XRD as a function of anneal time. The carbide grain sizes in (a) stay relatively constant during annealing, with any fluctuations being within the expected error (shown as a black error bar) of the XRD measurements. The carbide volume fractions in (b) also stay constant during annealing, with the exception of Cu-10Zr and Cu-11Hf which experience a slight decrease in carbide volume fraction during the early stages of annealing. (c) HAADF STEM micrograph and (d) EDS line scan from a Cu-4Hf sample annealed for one week, showing the presence of carbides with various sizes.

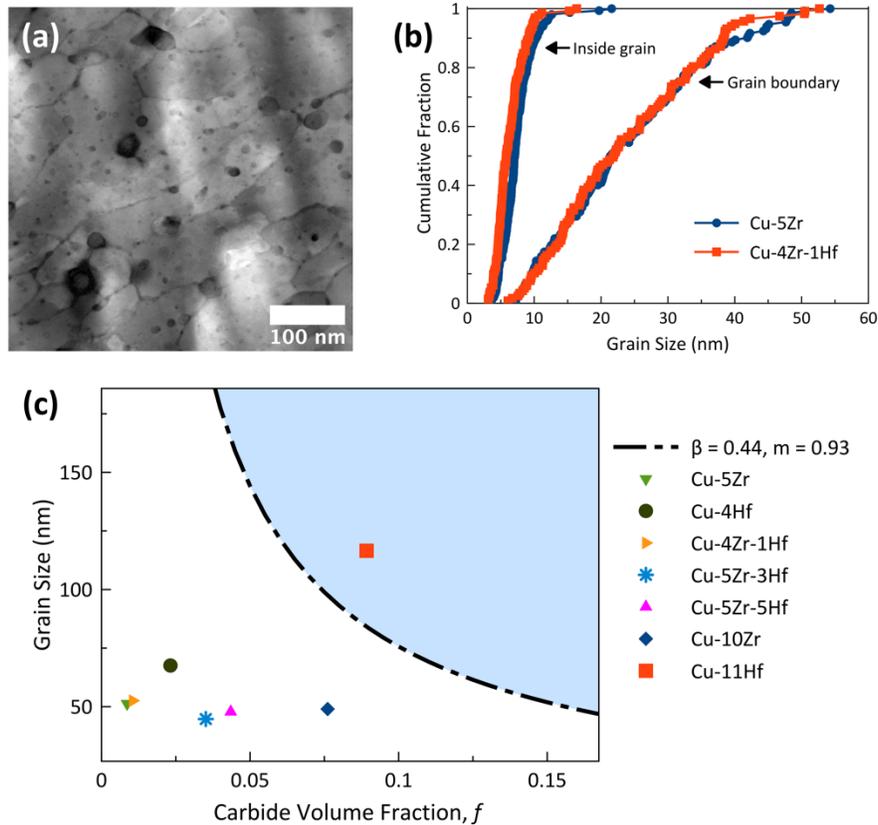

**Figure 5.** (a) HAADF STEM micrograph showing the distribution of carbides within grain interiors and at grain boundaries in the Cu-5Zr alloy annealed for 1 week at 950 °C. (b) Cumulative distribution functions demonstrating the similarity in carbide size distribution at both the grain boundaries and grain interiors for the Cu-5Zr and Cu-4Zr-1Hf alloys. (c) Grain size estimated by the Zener equation for an average second phase (carbide in this case) particle size of 20 nm, versus second phase volume fraction, $f$. The black dashed line represents the predicted grain size for Zener-pinned grains from Equation 2. TEM grain sizes for samples annealed for 1 week as a function of $f$ are plotted on this figure. With the exception of Cu-11Hf, all alloys have grain sizes much smaller than that predicted by the Zener equation, indicating that microstructural stabilization by boundary doping and AIF formation plays a significant role in their thermal stability.

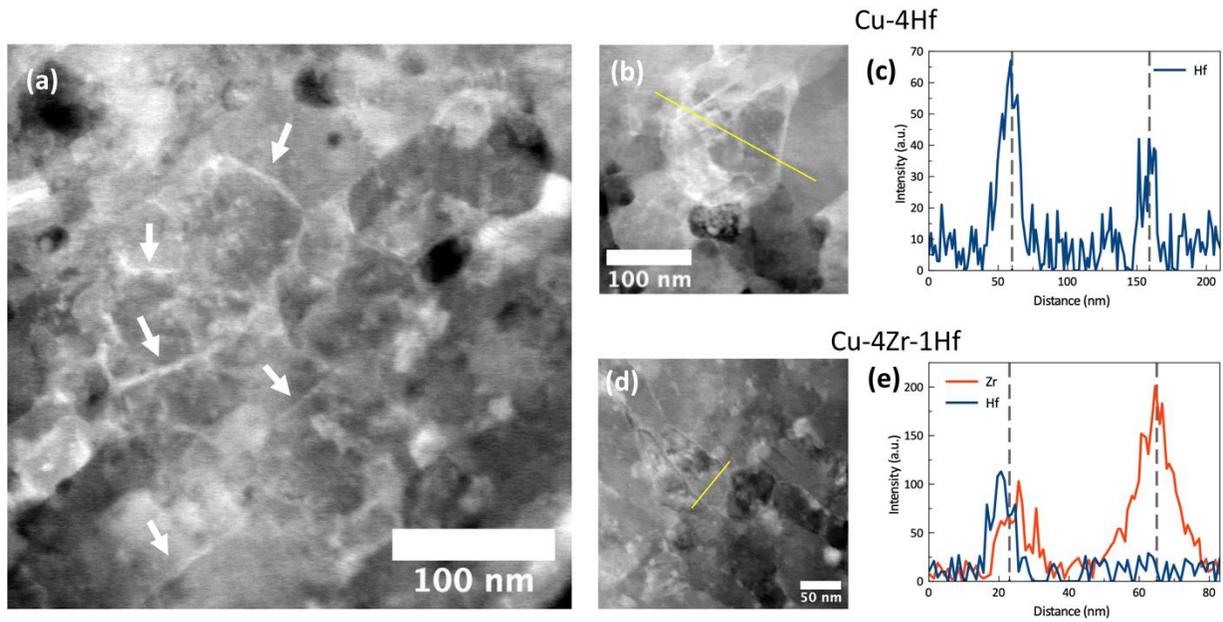

**Figure 6.** (a) HAADF STEM micrograph showing heavily doped boundaries in a Cu-4Zr-1Hf alloy annealed for 1 week. Arrows point to regions of brighter contrast associated with Zr and Hf segregation to these grain boundaries. (b) HAADF STEM micrograph and (c) corresponding EDS line scan showing Hf segregation to grain boundaries in a Cu-4Hf sample annealed for 1 week. (d) HAADF STEM micrograph and (e) corresponding EDS line scan showing segregation of both Zr and Hf to grain boundaries in a Cu-4Zr-1Hf sample annealed for 1 week. In both cases, the grain boundaries are heavily decorated with the dopants. The location of the grain boundaries in parts (c) and (e) are indicated with grey dotted lines.

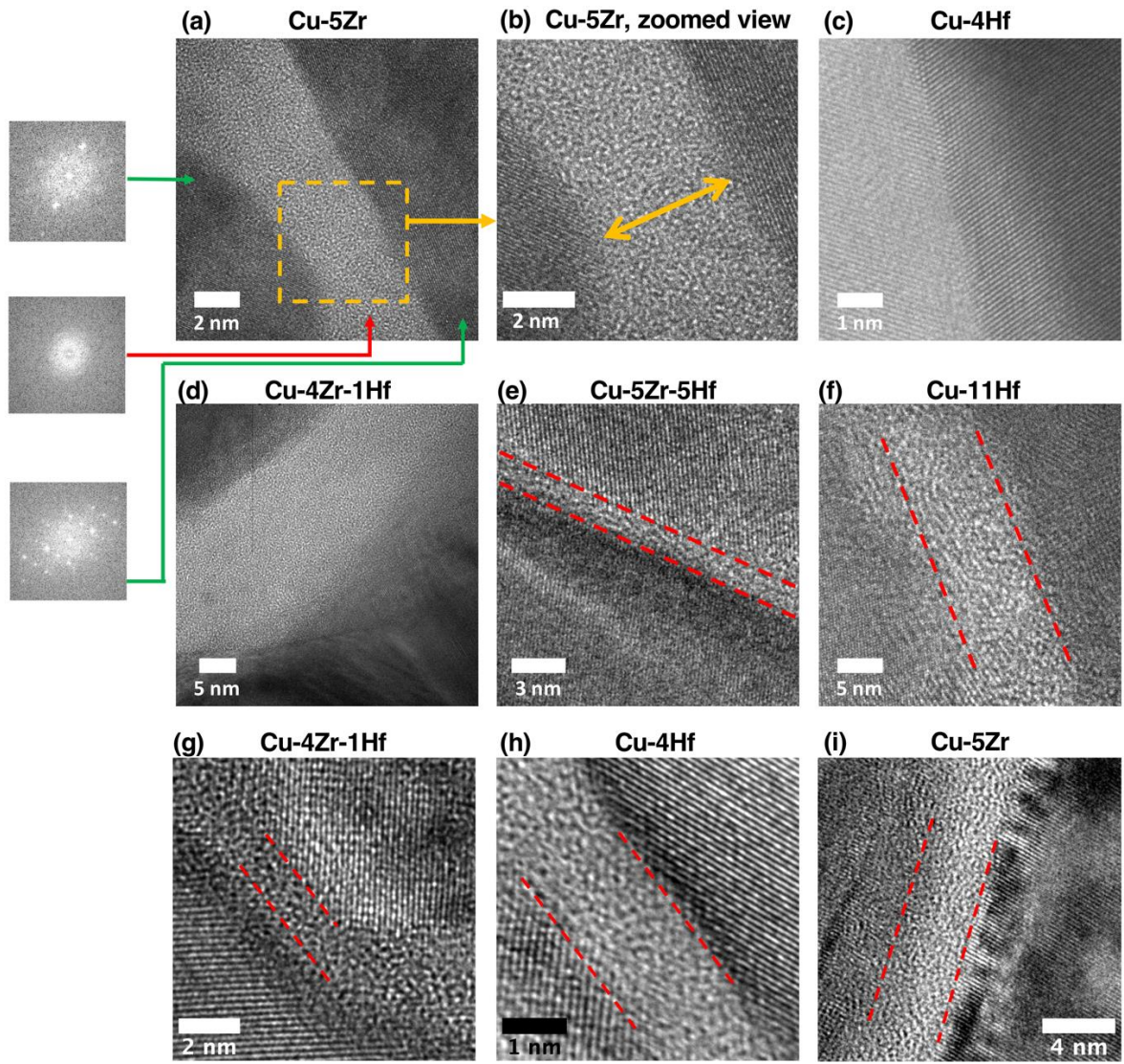

**Figure 7.** HRTEM micrographs of various complexions found in binary and ternary alloys annealed for 1 week. Dotted red lines denote AIFs between two crystalline regions. In (a), fast Fourier transforms of the crystalline and amorphous regions are shown on the left and the region that is edge-on is highlighted by a dashed yellow box. In (b), a zoomed view of the complexion in (a) to display the region of the grain boundary complexion where thickness measurements are made. The thickness of the complexion shown in (a) and (b) is 4.1 nm. A clean boundary in Cu-3Hf is shown in (c).

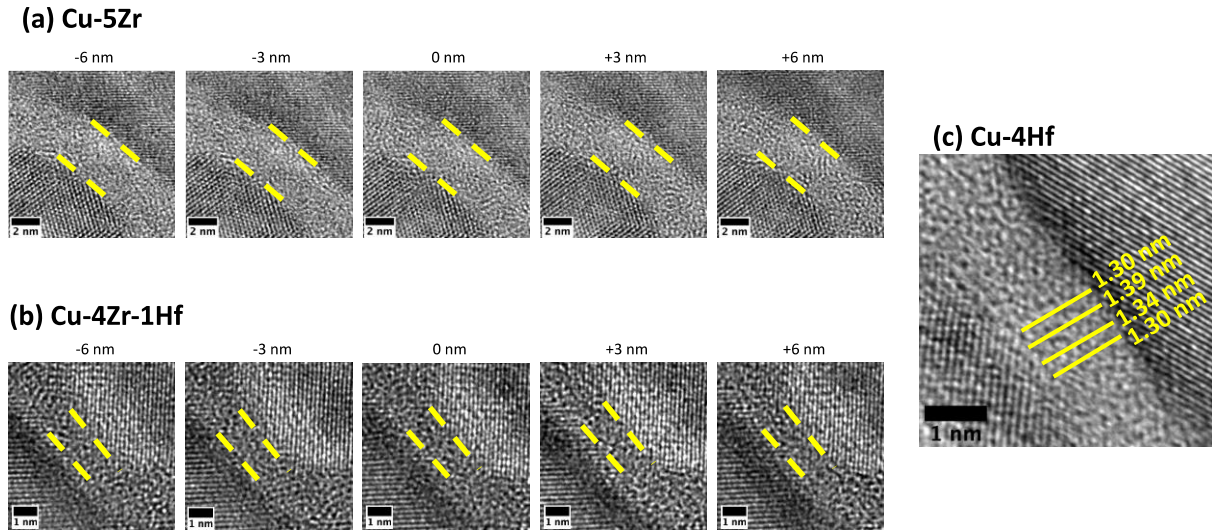

**Figure 8.** HRTEM micrograph of complexions in (a) a Cu-5Zr sample (scale bar 5 nm), (b) a Cu-4Zr-1Hf sample (scale bar 2 nm), and (c) a Cu-4Hf sample annealed for one week at 950 °C. The AIF thicknesses of 3.4 nm in the (a) Cu-5Zr sample and 1.2 nm in the (b) Cu-4Zr-1Hf sample remains unchanged in under-focused, focused, and over-focused imaging conditions, demonstrating that the grain boundary complexion is edge-on. (c) HRTEM image of an AIF in Cu-4Hf, showing the potential measurement error when taking care to measure from the thinnest region of the AIF. A maximum thickness variation of less than 0.1 nm is found.

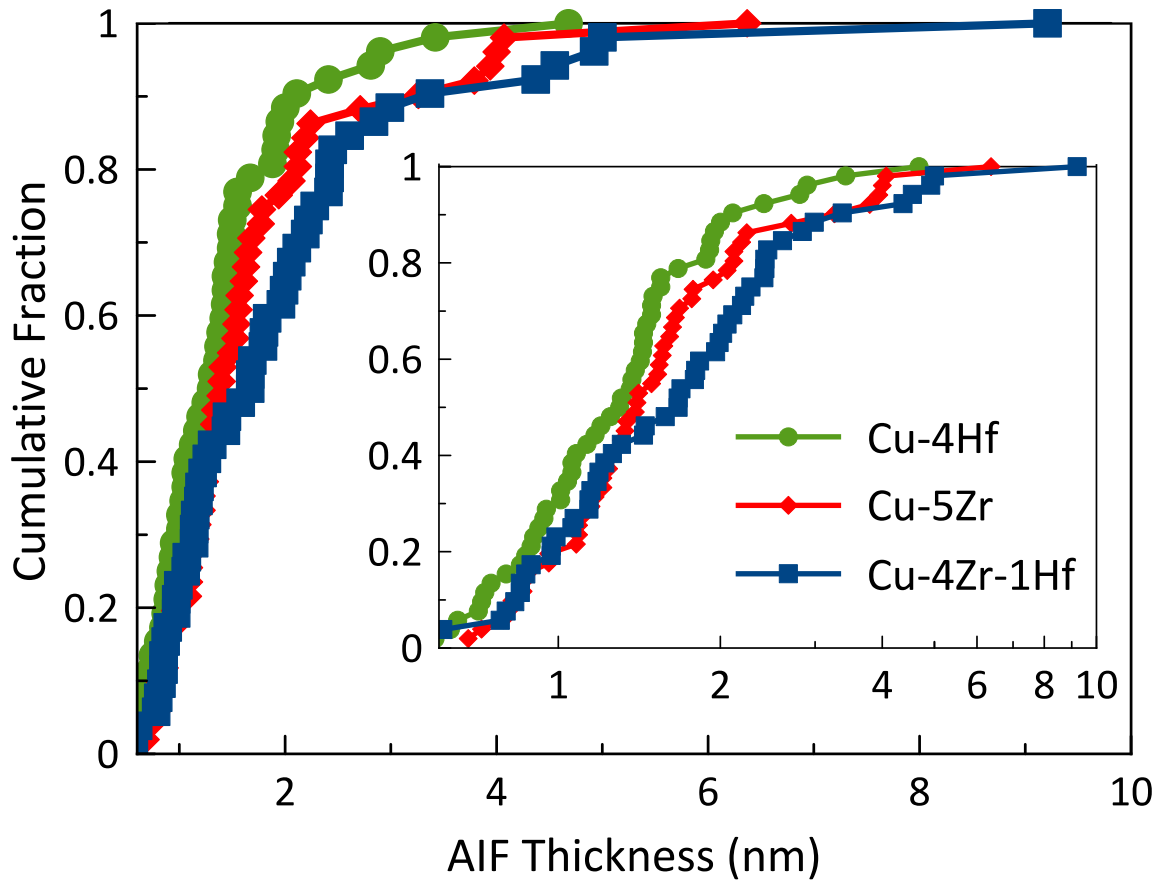

**Figure 9.** Cumulative distribution function of AIF thicknesses in Cu-5Zr, Cu-4Hf, and Cu-4Zr-1Hf alloys annealed for 1 week. At least 50 AIFs were measured from HRTEM micrographs of grain boundaries in an edge-on condition in each of the three samples. Distributions pushed to the right indicate thicker AIFs. Complexions in the Cu-4Zr-1Hf alloy are noticeably thicker than those found in binary alloys.
36